\documentclass[aps,xpre,longbibliography,showpacs,amsmath,amssymb]{revtex4-1}
\usepackage{amsmath,amssymb,latexsym,epsfig,graphics,epsf}
\usepackage{hyperref}
\usepackage{graphics}
\usepackage{float}
\usepackage{footnote}
\usepackage[utf8]{inputenc}
\usepackage{adjustbox}
\usepackage{blindtext}
\usepackage[T1]{fontenc}
\usepackage{lpic}
\usepackage{tabularx,booktabs}
\usepackage[dvipsnames]{xcolor}
\usepackage{ulem}

\newcommand{\be}{\begin{equation}}
\newcommand{\ee}{\end{equation}}
\newcommand{\fig}[1]{Fig.~\ref{#1}}
\newcommand{\figs}[1]{Figs.~\ref{#1}}
\newcommand{\Fig}[1]{Figure~\ref{#1}}
\newcommand{\Figs}[1]{Figures~\ref{#1}}
\newcommand{\eq}[1]{Eq.~(\ref{#1})}
\newcommand{\Eq}[1]{Equation~(\ref{#1})}

\newcommand{\Tc}{{T}_{\rm conf}}

\newcommand{\bR}{\mathbf{R}}
\newcommand{\bRa}{{\bf R}_{\rm a}}
\newcommand{\bRb}{{\bf R}_{\rm b}}
\newcommand{\br}{\mathbf{r}}

\begin{document}
	\title{\texorpdfstring{$NVU$}{NVU} view on energy polydisperse Lennard-Jones systems} 
	\date{\today}
        \author{Danqi Lang}
        \author{Lorenzo Costigliola}
	\author{Jeppe C. Dyre}\email{dyre@ruc.dk}
	\affiliation{\textit{Glass and Time}, IMFUFA, Department of Science and Environment, Roskilde University, P.O. Box 260, DK-4000 Roskilde, Denmark}
	
\begin{abstract}
When energy polydispersity is introduced into the Lennard-Jones (LJ) system, there is little effect on structure and dynamics [Ingebrigtsen and Dyre, J. Phys. Chem. B \textbf{127}, 2837 (2023)]. For instance, at a given state point both the radial distribution function and the mean-square displacement as a function of time are virtually unaffected by even large energy polydispersity, which is in stark contrast to what happens when size polydispersity is introduced. We here argue -- and validate by simulations of up to 30\% polydispersity -- that this almost invariance of structure and dynamics reflects an approximate invariance of the constant-potential-energy surface. Because $NVU$ dynamics defined as geodesic motion at constant potential energy is equivalent to Newtonian dynamics in the thermodynamic limit, the approximate invariance of the constant-potential-energy surface implies virtually the same physics of energy polydisperse LJ systems as of the standard single-component version. In contrast, the constant-potential-energy surface is significantly affected by introducing size polydispersity.
\end{abstract}
\maketitle

\vspace{1cm}

\section{Introduction}

Polydispersity typically involves a continuous distribution of parameters in the interaction potentials of models, e.g., of liquids. This is relevant for describing the jamming of granular media, which can be modeled by introducing a distribution of particle sizes \cite{dickinson1978,salacuse1982,fre86,evans1999,weeks2000,fasolo2003,abraham2008,ballesta2008,nguyen2014,ing15}. Size polydispersity is also relevant in the modeling of glass-forming liquids by allowing for fast equilibration via swap dynamics; here it is typically introduced via a distribution of the length parameter of the pair potential \cite{nin17}. If the ``size'' of particle $i$ is $\sigma_i$, the Lorentz-Berthelot mixing rule states that the interaction with particle $j$ involves the average length parameter, $(\sigma_i+\sigma_j)/2$ \cite{tildesley}. While \textit{size} polydispersity is by far the most commonly studied sort of polydispersity \cite{dickinson1978,salacuse1982,gualtieri1982,kofke1986,evans1999,auer2001,fasolo2003,wilding2005,abraham2008,ballesta2008,jacobs2013,sar13,nguyen2014}, a few publications have investigated the effects of energy polydispersity \cite{shagolsem2015a,ing16,sha16,ing23}. Using the Lorentz-Berthelot mixing rule for energies \cite{tildesley}, Refs. \onlinecite{shagolsem2015a} and \onlinecite{sha16} studied energy-polydisperse Lennard-Jones (LJ) fluids in 2d and found only very small differences between the average properties of polydisperse 
systems and those of the single-component fluid with interaction energy equal to the average of that of the polydisperse system. This was recently confirmed in a study of 3d energy-polydisperse LJ mixtures, demonstrating virtually invariant structure and dynamics when the degree of polydispersity varies \cite{ing23}. Energy-polydispersity invariance is robust; thus Ref. \onlinecite{ing23} demonstrated that -- excluding the case of extreme energy polydispersity -- invariance is maintained when varying state point, mixing rule, energy probability distribution, or the pair potential. The question we address in this paper is: Why does the introduction of energy polydispersity not affect the structure and dynamics to any significant degree?

In order to throw light on the energy-polydispersity invariance of structure and dynamics, we adopt below the $NVU$ point of view. In a sense, $NVU$ dynamics replaces Newton's second law by his first law -- the law of inertia -- by considering geodesic motion on the constant-potential-energy hypersurface in $3N$ dimensions ($N$ is the number of particles) \cite{NVU_I,NVU_II}. In the thermodynamic limit ($N\to\infty$) this time evolution results in the same structure and dynamics as standard Newtonian dynamics \cite{NVU_I,NVU_II}. In fact, if $NVU$ dynamics is discretized for numerical implementation, the result is identical to the well-known Verlet algorithm \cite{ver68,tildesley,han13} with a varying time step, the relative fluctuations of which go to zero for $N\to\infty$ \cite{NVU_II}. The conjecture investigated in this paper is that energy-polydisperse LJ mixtures have approximately the same constant-potential-energy surfaces as those of the single-component LJ system, which would explain the observed invariance of structure and dynamics upon the introduction of energy polydispersity.

\section{Approximate invariance of structure and dynamics}

This section details the systems studied, how they were simulated, and gives results for the structure and dynamics confirming those of Ref. \onlinecite{ing23}. We simulated energy-polydisperse Lennard-Jones (LJ) liquids in the $NVT$ ensemble using a pilot version of an in-house developed GPU-optimized Python Molecular Dynamics code \cite{rumdpy24}. The LJ pair potential between two particles at distance $r$, $v(r)$, is given by 

\be\label{LJ_pot}
v(r)
\,=\,4\varepsilon \left((r/\sigma)^{-12}-(r/\sigma)^{-6}\right)\,.
\ee
Here $\varepsilon$ is the characteristic energy and $\sigma$ the characteristic length (``particle size''). For polydisperse systems these parameters vary for different particle pairs, usually according to a continuous probability distribution. This paper focuses on energy polydispersity. Thus $\sigma$ is the same for all particles unless otherwise stated, henceforth set to unity, whereas the characteristic energy of the $ij$ pair interaction follows the Lorentz-Berthelot mixing rule \cite{tildesley}, $\varepsilon_{ij}=\sqrt{\varepsilon_i\varepsilon_j}$, in which each particle has been assigned an energy $\varepsilon_i$ chosen randomly from a box distribution centered around unity (in practice we operated with 256 different equally spaced randomly chosen particle energies). The energy polydispersity $\delta_\varepsilon$ is defined as the standard deviation over the mean, which in the case of a box distribution with unit mean energy reduces to $\delta_\varepsilon\equiv \sqrt{\langle\varepsilon^2\rangle-1}$.

\begin{figure}[htbp]
    \centering
    \includegraphics[width=0.6\linewidth]{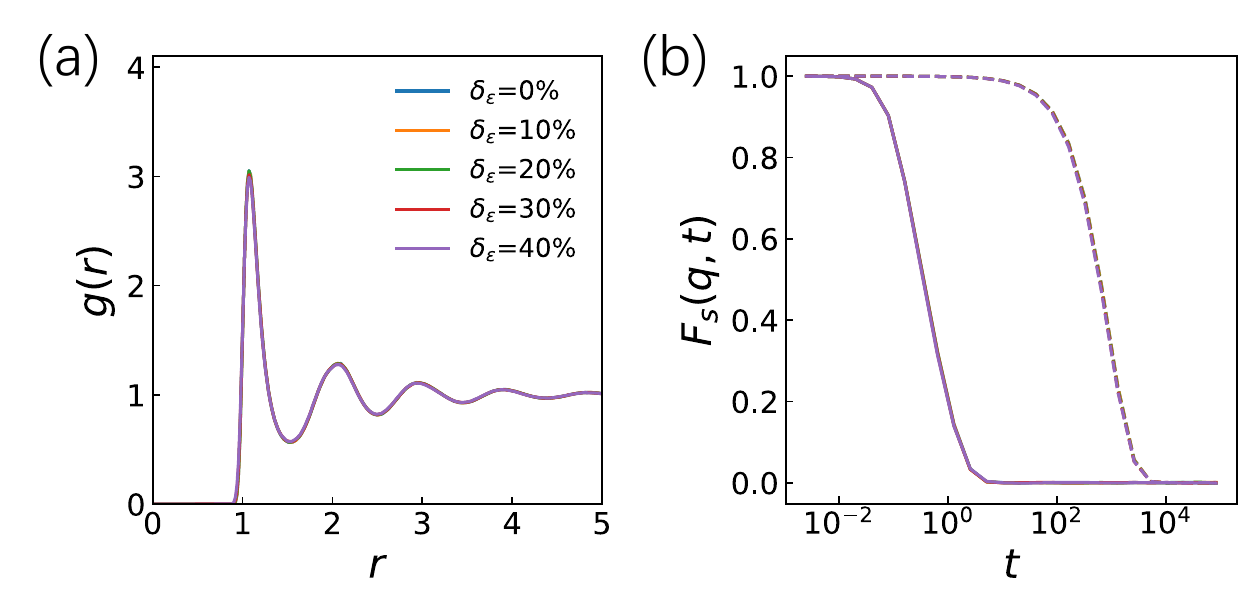}
    \caption{Average structure and dynamics of LJ systems of different energy polydispersity at the state point $\left(\rho,T\right) = (0.85,0.70)$ that is close to the triple point of the single-component LJ system. 
    (a) shows the average radial distribution function, $g(r)$, for up to 40\% polydispersity. 
    (b) shows the corresponding average incoherent intermediate scattering functions, $F_s(q,t)$, where the five solid lines represent data for the wave vector of the first peak of the static structure factor of the monodisperse system ($q = 7.2$) and the five dashed lines represent data for the wave vector corresponding to the box length ($q = 0.19$). 
    Structure and dynamics are virtually independent of the degree of polydispersity. These results confirm those of Ref. \onlinecite{ing23} and set the stage for the investigation.}
    \label{fig1}
\end{figure}

Systems of $N=32000$ particles of mass unity were simulated with standard Nose-Hoover $NVT$ dynamics at the state point $(\rho, T) = (0.85, 0.70)$, which is close to the triple point of the single-component LJ system. This is the state point studied throughout the paper. The time step used was $0.0025$ in LJ units (defined by $\varepsilon=\sigma=1$ in \eq{LJ_pot}). A standard shifted-potential cutoff at $2.5$ was used. We also simulated LJ systems with a shifted-force cutoff at 1.5 \cite{tox11a}, leading to the same overall conclusions (Appendix).

\begin{figure}
    \centering
    \includegraphics[width=1.0\linewidth]{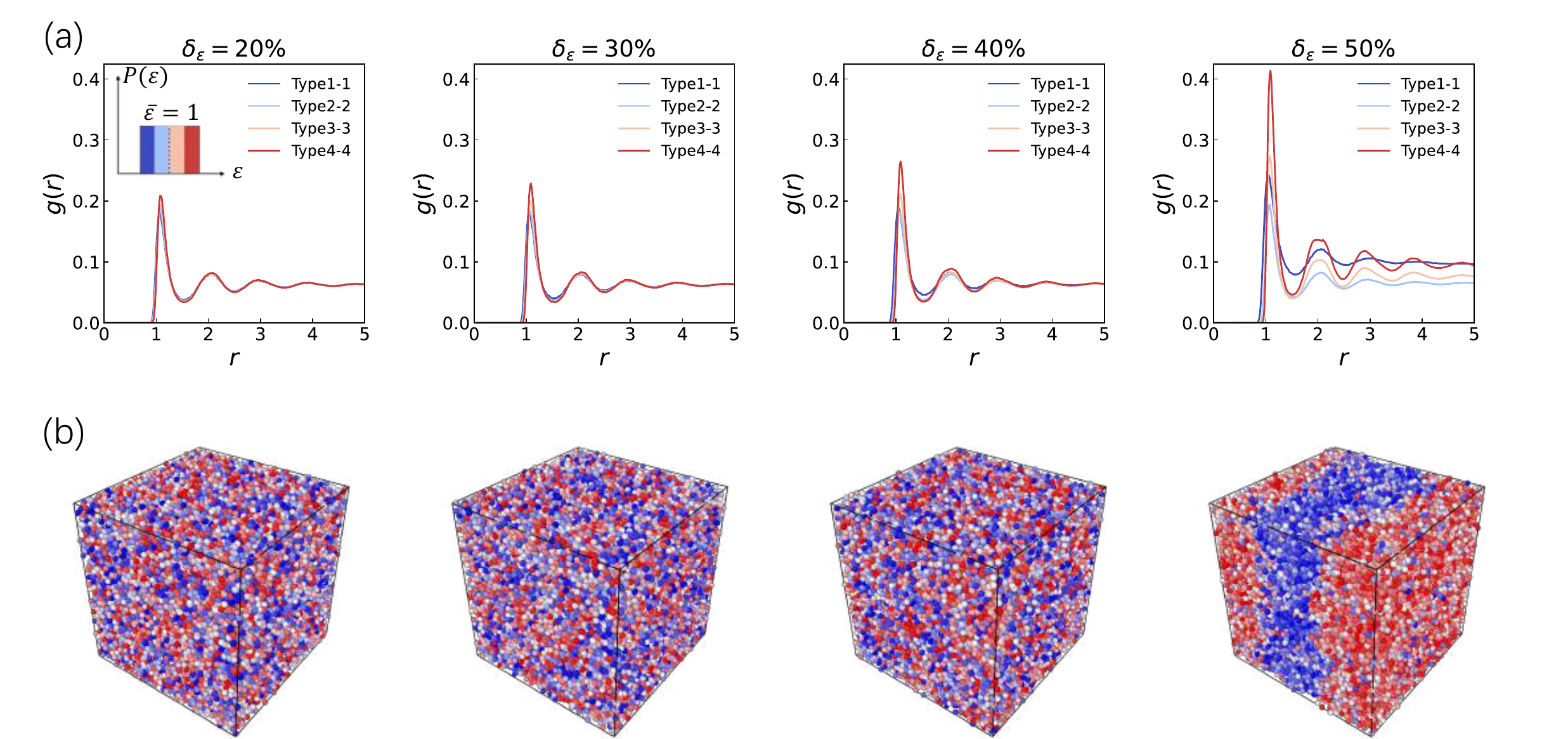}
    \caption{Radial distribution functions (RDFs) of LJ systems with energy polydispersity varying between 20\% and 50\%. 
    (a) For each polydispersity the particles are divided into four categories according to their energy as illustrated in the inset of the left of panel (a). With increasing polydispersity the energy-resolved RDFs differ more and more.
    (b) shows snapshots for each polydispersity in which the red particles have large energy and the blue have small energy. }
    \label{fig2}
\end{figure}

\Fig{fig1} shows the structure and dynamics for energy polydispersity $\delta_\varepsilon=0\%$; $10\%$; $20\%$; $30\%$; $40\%$. The first case corresponds to the standard single-component LJ system. The next four cases correspond to particle energies varying between 0.83 and 1.17 (10\% polydispersity), between 0.65 and 1.25 (20\% polydispersity), between 0.48 and 1.52 (30\% polydispersity), and between 0.31 and 1.69 (40\% polydispersity). These are quite significant polydispersities. Nevertheless, the average radial distribution function (RDF), $g(r)$, of \fig{fig1}(a) and the average incoherent intermediate scattering function, $F_s(q,t)$, of \fig{fig1}(b) are virtually the same for all five systems. 

\Fig{fig2} analyzes to which degree the neighborhood of a given particle correlates with its energy for up to 50\% polydispersity. We divided the particles into four categories according to their energy, each of which contains one quarter of the particles. \Fig{fig2}(a) shows that as polydispersity is increased, the energy-resolved RDFs differ increasingly, i.e., the neighborhood of each particle depends more and more on its energy. As shown in the right panel of \fig{fig2}(b), the case of 50\% polydispersity self-organizes by a continuous phase separation into high (red) and low (blue) energy particles \cite{ing23}.

\begin{figure}
    \centering
    \includegraphics[width=0.5\linewidth]{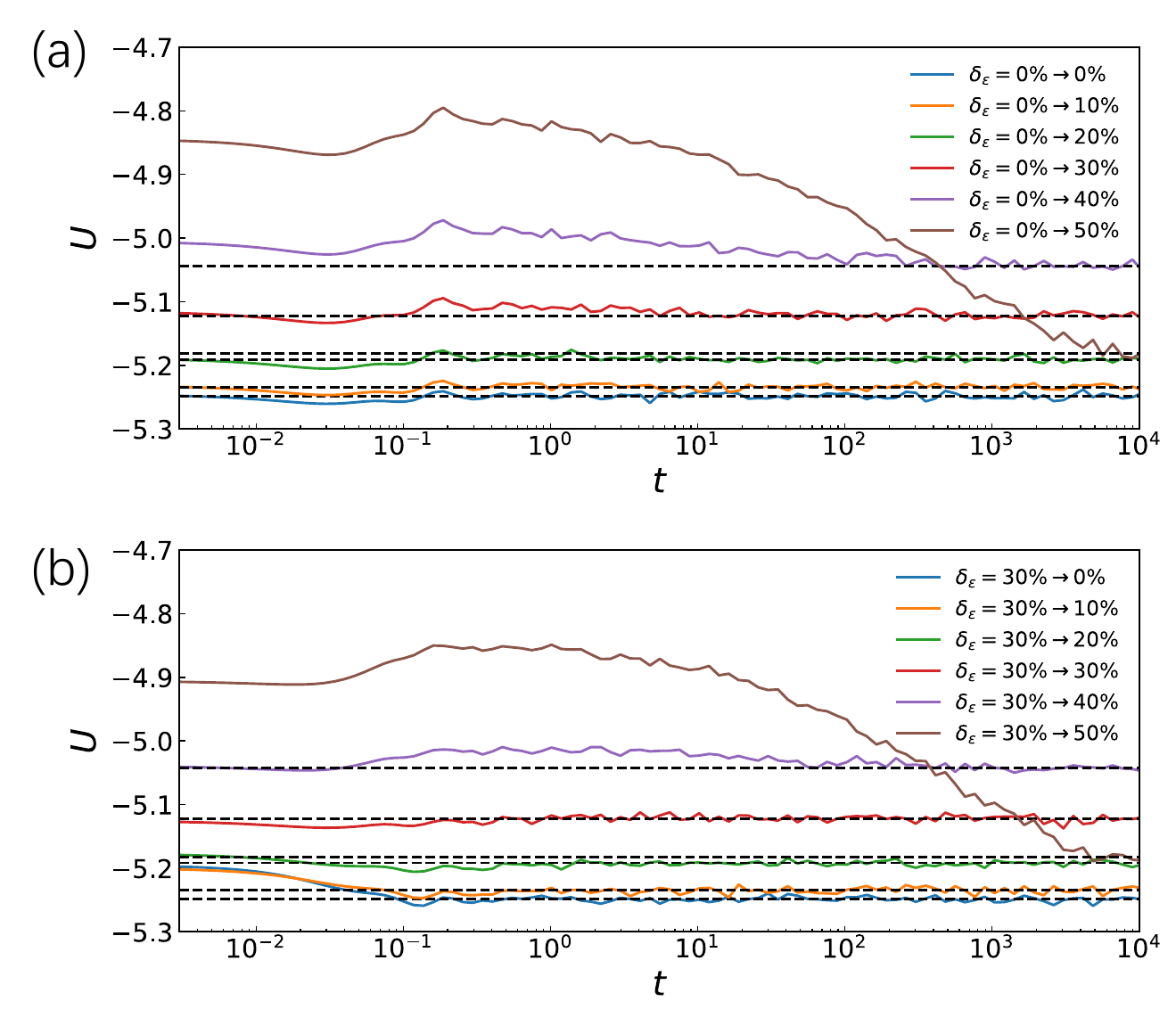}
    \caption{Potential-energy relaxation after an instantaneous change of energy polydispersity. The dashed lines mark the equilibrium potential energy approached at long times. 
    (a) Jumping from the equilibrium single-component system to different polydispersities; 
    (b) jumping from an equilibrium 30\% polydispersity system to different polydispersities.
    In most cases little happens, which is consistent with the system being in equilibrium already immediately after the polydispersity was changed at $t=0$. For jumps to 50\% polydispersity this is not the case, however; the 50\% data confirm the phase separation observed as this system slowly equilibrates (\fig{fig2}(b)). For the jumps to 40\% polydispersity there is a slight tendency of a similar behavior. }
    \label{fig5}
\end{figure}

To illuminate the self-organization observed for 50\% polydispersity, we study in \fig{fig5} how the potential energy relaxes toward equilibrium after polydispersity is introduced. The idea is that if there is phase separation deriving from self-organization, the energy will relax slowly toward its equilibrium value (a process that is controlled by the particle diffusion coefficient and the simulation box size). \Fig{fig5}(a) shows the time evolution of the average potential energy per particle, $U$, after jumps at $t=0$ from the single-component system to polydispersities up to 50\%, and (b) shows similar data for jumps starting from 30\% polydispersity. In both cases the jumps to and from up to 30\% polydispersity equilibrate quickly. In contrast, jumps to 50\% polydispersity show a slow relaxation toward equilibrium, confirming that self-organization takes place in this case. The jumps to 40\% show a slight hint of the same behavior. Henceforth we leave out 40\% and 50\% polydispersity from the analysis.

\section{Digression: \texorpdfstring{$NVU$}{NVU} dynamics in brief}

This section briefly reviews $NVU$ dynamics \cite{NVU_I,NVU_II,NVU_III,dyr13,dyr13a} for readers unfamiliar with this alternative molecular dynamics characterized by strict conservation of the potential energy. For systems of many particles, Newtonian and $NVU$ dynamics lead to the same structure and dynamics \cite{NVU_II}. This may be understood as a consequence of the fact that in the thermodynamic limit of Newtonian dynamics, the \textit{relative} fluctuations of the potential energy $U$ go to zero. Identical structure and dynamics has been shown theoretically and numerically for both atomic \cite{NVU_I,NVU_II} and molecular \cite{NVU_III} models. In fact, as already mentioned the discretized $NVU$ algorithm is identical to the standard Verlet algorithm of molecular dynamics simulations, however with a varying time step ensuring conservation of the potential energy \cite{NVU_I}.

Consider a system of $N$ particles in three dimensions with periodic boundary conditions. It is convenient to introduce the $3N$-dimensional configuration vector $\bR\equiv (\br_1,...,\br_N)$ in which $\br_i$ is the position of particle $i$. A constant-potential-energy surface $\Omega$ is defined by the value of the potential energy $U_0$, 

\be\label{Omega_def}
\Omega
\,=\,\{\bR\,|\,U(\bR)=U_0\}\,.
\ee
\Eq{Omega_def} defines a $3N-1$ dimensional so-called level surface, which is a submanifold of the $3N$-dimensional torus of all particle positions corresponding to periodic boundary conditions. The surface $\Omega$ has the metric inherited from Euclidean space and is thus a Riemannian manifold. This makes it possible to define \textit{geodesics} as curves of minimum length or, more accurately, of stationary length in the sense that the length does not change for infinitesimal curve perturbations keeping the endpoints fixed. $NVU$ dynamics is defined as geodesic motion on $\Omega$ \cite{NVU_I}. The motion proceeds with constant velocity, implying that, in fact, both the potential and the kinetic energy are conserved in $NVU$ dynamics. In physical terms, $NVU$ dynamics may be thought of as embodying friction-less motion on a curved surface in high dimensions, i.e., as expressing the law of inertia for motion on the curved hypersurface $\Omega$. 

It follows that systems with identical constant-potential-energy surfaces have same structure and dynamics, also when studied by standard Newtonian dynamics. In many cases, the same $\Omega$ at given state points of two systems corresponds to different temperatures, but this turns out to be a minor effect in this study that can be ignored. We investigate below whether the $\Omega$ surfaces change only little when energy polydispersity is introduced.

\begin{figure}
    \centering
    \includegraphics[width=0.8\linewidth]{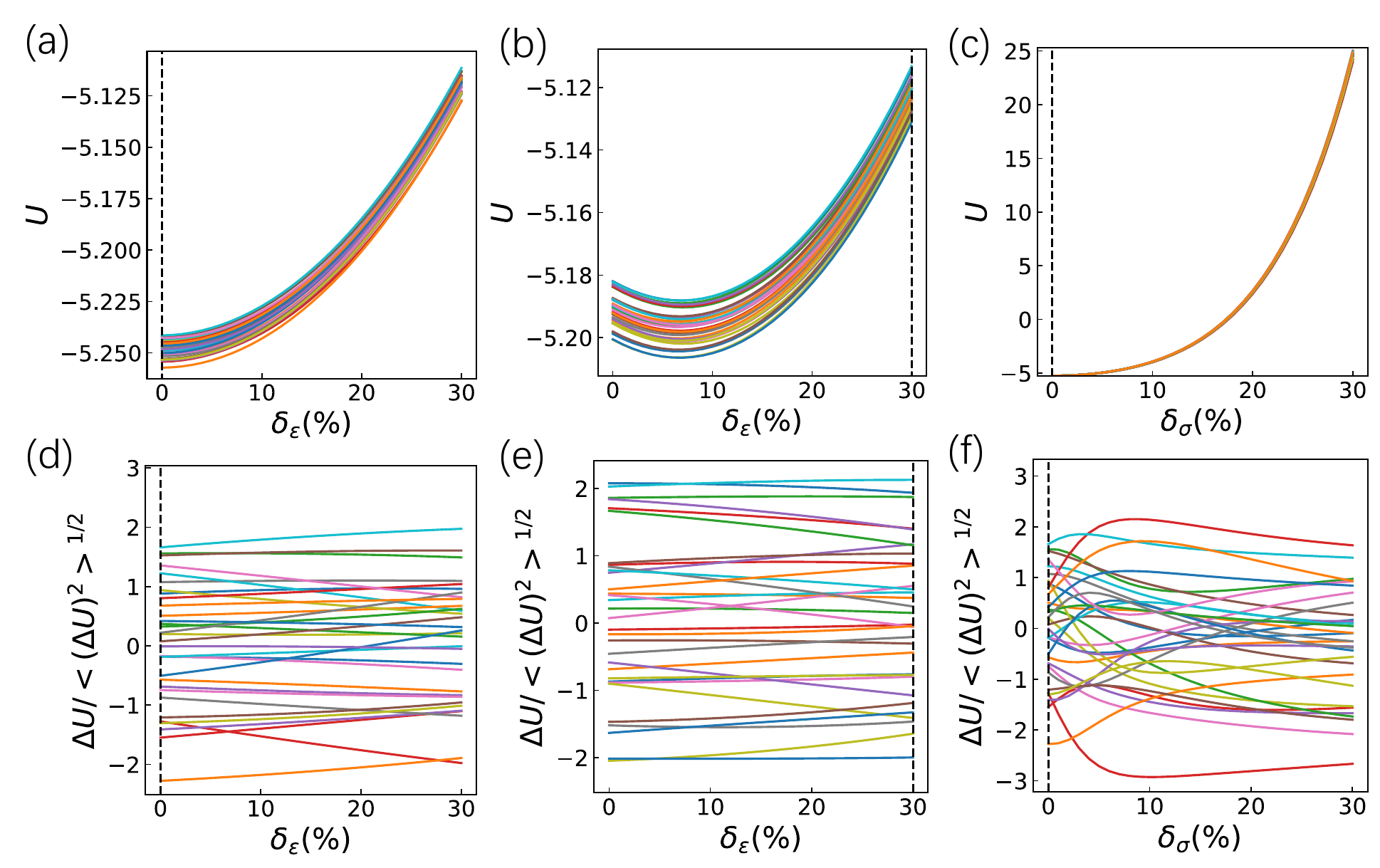}
    \caption{Potential energies of configurations as a function of the degree of energy polydispersity. 
    (a) and (b) show the potential energies of 32 independent configurations taken from an equilibrium simulation at 0\% and 30\% energy polydispersity, respectively. Once the configurations have been selected at the polydispersity marked by the vertical dashed line, the degree of polydispersity is varied continuously in the expression for $U(\bR)$, i.e., no further simulations are carried out. A closer look at the crossings is provided in (d) and (e), which show the relative potential-energy variations of the data of (a) and (b). The observed rare crossings mean that energy polydisperse LJ systems to a good approximation conform to \eq{crit}, i.e., have invariant $\Omega$. For comparison, (c)  and (f) show the effect of introducing size polydispersity. }
    \label{fig3}
\end{figure}

\section{Approximate invariance of the constant-potential energy surface $\Omega$}

\begin{figure}
    \centering
    \includegraphics[width=\linewidth]{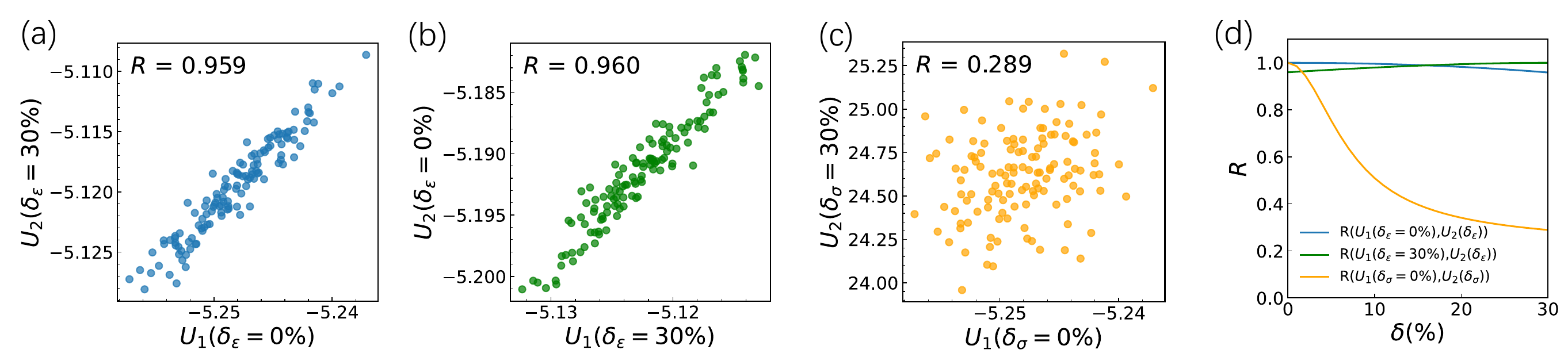}
    \caption{Correlations between the potential energy of configurations sampled from the equilibrium state of potential $U_1$ corresponding to some polydispersity, subsequently evaluated with respect to a different polydispersity, corresponding to the potential energy $U_2$. 
    (a) Correlation between the single-component system, $U_1$, and 30\% energy polydispersity, $U_2$. 
    (b) The reverse of (a); configurations were here selected from a 30\% energy polydispersity simulation and subsequently evaluated with respect to zero polydispersity. The strong correlations observed in (a) and (b) are in contrast to what happens for size polydispersity; thus (c) shows the analogous correlation between the potential energies of the single-component system and that of 30\% size polydispersity. (d) shows how the Pearson correlation coefficient $R$ varies as a function of polydispersity for the two cases of energy polydispersity (blue and green corresponding to (a) and (b)), and for size polydispersity (yellow). In the latter case the correlation coefficient drops quickly.}
    \label{fig4}
\end{figure}

Returning to the invariance to a very good approximation of structure and dynamics when energy polydispersity is introduced, the observations of \fig{fig1} would as mentioned be explained if the constant-potential-energy surface $\Omega$ is independent -- or almost independent -- of the energy polydispersity. We proceed to investigate whether this is the case. Note first, however, that there is no theoretical reason a given $\Omega$ is completely unaffected by the degree of energy polydispersity: any two configurations that have identical potential energy at one polydispersity, will most likely have different potential energies for a different polydispersity. Checking invariance of the identity $U(\bRa)=U(\bRb)$ is therefore not useful for investigating approximate invariance of the constant-potential-energy surface. Instead, we take inspiration from isomorph theory that considers invariance of the inequality $U(\bRa)<U(\bRb)$ when configurations are scaled uniformly \cite{sch14}. Comparing two potential-energy functions corresponding to different energy polydispersity, $U_1(\bR)\equiv U(\bR,\delta_\varepsilon^{(1)})$ and $U_2(\bR)\equiv U(\bR,\delta_\varepsilon^{(2)})$, we investigate numerically how well the following logical implication is obeyed

\be\label{crit}
U_1(\bRa)<U_1(\bRb)\,\implies\ U_2(\bRa)<U_2(\bRb)\,.
\ee
If this applies rigorously for all configurations, then $U_1(\bRa)=U_1(\bRb)\,\implies\ U_2(\bRa)=U_2(\bRb)$, i.e., systems $1$ and $2$ have identical $\Omega$. If however one more realistically observes that \eq{crit} applies for most configurations, though not always, then the corresponding constant-potential-energy surfaces are not identical, but merely almost identical. In that case, by reference to $NVU$ dynamics one expects approximately, not rigorously invariant structure and dynamics.

To check \eq{crit} numerically we first sampled several configurations from an equilibrium simulation of system 1, e.g., the single-component LJ system. The configurations were sampled at times separated enough that they are statistically independent. For each configuration, we then evaluated the potential energy corresponding to a different degree of polydispersity in order to investigate how this quantity changes, i.e., no further MD simulations were carried out. Plotting the potential energies as functions of the polydispersity yields a figure confirming \eq{crit} if none of the curves cross each other. 

\Fig{fig3}(a) shows a plot constructed in this way in which the configurations were selected from an equilibrium simulation of the single-component system; (b) shows the analog if configurations are selected from an equilibrium simulation at 30\% energy polydispersity (in which case the lower polydispersities were obtained by a uniform scaling of the energies relative to unity). One first observes that the value of the potential energy changes only a little. Moreover, there are only few curve crossings, which largely validates \eq{crit}. \Figs{fig3}(d) and (e) study the crossings in more detail by plotting the relative variation of the potential energy as a function of the degree of polydispersity, i.e., after the average of $U$ has been subtracted at each polydispersity and the data subsequently normalized to unit variance. The vertical dashed lines mark the polydispersity of the system simulated to generate the configurations. For comparison, \fig{fig3}(c) and (f) show the same when size polydispersity is introduced. While for energy polydispersity there are only few crossings, and when two curves do cross they generally stay close, neither of these observations apply for size polydispersity. 

Note that there is little difference between selecting the configurations from the single-component LJ system and then introducing polydispersity ((a) and (c)) and the alternative of selecting the configurations at 30\% energy polydispersity and subsequently decreasing the polydispersity ((b) and (d)). This shows that the configurations remain equilibrium configurations to a good approximation after changing the polydispersity.

The results shown in \figs{fig3}(a), (b), (d), and (e) refer to one particular choice of random energies. We have repeated this analysis for several other choices and -- reflecting the fact that large samples are studied -- have found that the results are the same. Thus for different choices of random energies or different choices of equilibrate configurations, \figs{fig3}(a) and (b) are basically unchanged, while \figs{fig3}(d) and (e) can be visually different, but are always qualitatively the same with few level crossings (independence of configurations and random parameters applies also to the size-polydispersity results of \fig{fig3}(c) and (f)). \Figs{sfig_U_otherrandomseed} - \ref{sfig_RforU_otherrandomseed_30to0} in the Appendix demonstrate that different choices of polydispersity qualitatively lead to the same conclusions.

\Fig{fig3} features several instances of weak ``level'' crossings in the sense that the two potential energies stay close even if they cross at some polydispersity. Intuitively, one expects this implies a less severe violation of the proposed invariance of $\Omega$ than if, e.g., the energies vary wildly as in the case of size polydispersity (\fig{fig3}(f)). How to quantify this? A simple possibility is to calculate how the Pearson correlation coefficient $R$ between the initial and final potential energies, $U_1$ and $U_2$, varies as a function of the degree of polydispersity. \Fig{fig4}(a) shows a plot of $\delta_\varepsilon =0$\% data versus $\delta_\varepsilon =$30\% data, and \fig{fig4}(b) shows the reverse correlation plot. In both cases there is a strong correlation with $R>0.95$ (note that these plots involve many more configurations than the 32 of \fig{fig3}). This confirms the conjecture that energy polydispersity leads to only minor modifications of the constant-potential-energy surfaces.

It is instructive to compare to size polydispersity. This is studied in an analogous way in \fig{fig4}(c) starting from the monodisperse LJ system. This time there is a very poor correlation between the $\delta_\sigma =0$\% polydispersity and $\delta_\sigma =$ 30\% polydispersity. In fact, the correlation coefficient drops quickly as a function of polydispersity; this is clear from \fig{fig4}(d) that shows the correlation coefficients as a function of polydispersity for all three cases.

\begin{figure}
    \centering
    \includegraphics[width=\linewidth]{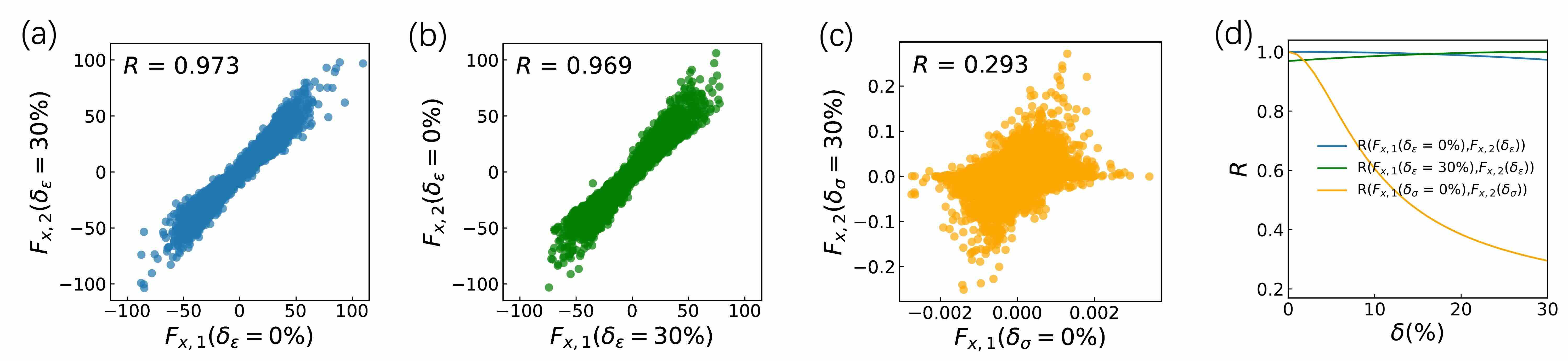}
    \caption{Correlation between the individual particle forces in the x-direction, $F_x$, of a single configuration sampled from the equilibrium state of some polydispersity, subsequently evaluated with respect to a different polydispersity. 
    (a) Correlation between the single-component system and 30\% energy polydispersity. 
    (b) The reverse of (a); configurations were here equilibrated at 30\% energy polydispersity and the forces were subsequently evaluated with respect to 0\% polydispersity. In both (a) and (b) there is a strong correlation. This is in contrast to what happens in the case of size polydispersity; thus 
    (c) shows a scatter plot between the single-component system and 30\% size polydispersity.
    (d) shows how the correlation coefficient $R$ varies as a function of polydispersity for the two cases of energy polydispersity (blue and green corresponding to (a) and (b)) and size polydispersity (yellow). In the latter case the correlation coefficient drops quickly.}
    \label{fig6}
\end{figure}

\section{Two further approximate invariants}

\begin{figure}
    \centering
    \includegraphics[width=\linewidth]{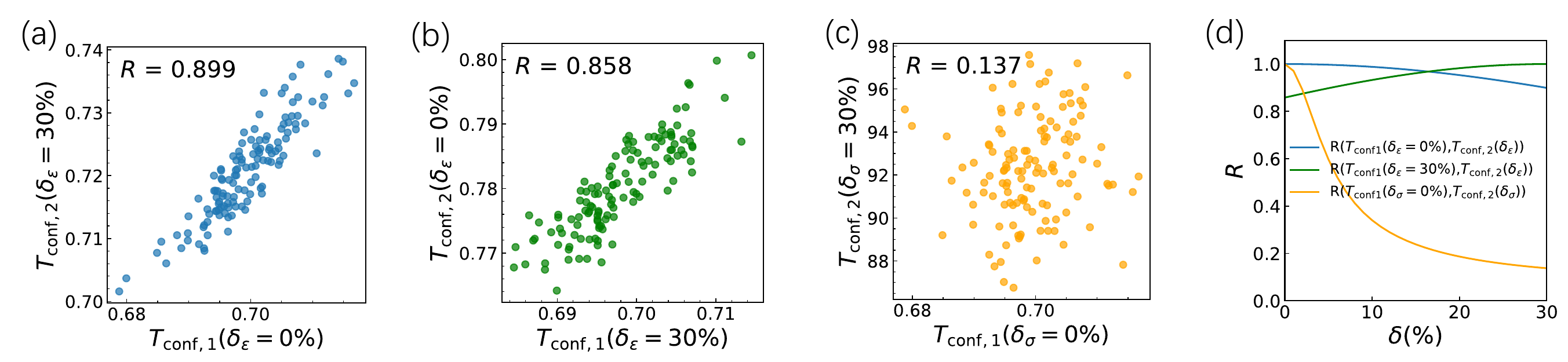}
    \caption{Correlation between the configurational temperatures, $\Tc$, of configurations sampled from the equilibrium state of some polydispersity and $\Tc$ evaluated with respect to a different polydispersity. 
    (a) Correlation between the single-component system and 30\% energy polydispersity. 
    (b) The reverse of (a); configurations were here equilibrated at 30\% energy polydispersity and the configurational temperatures were subsequently evaluated with respect to 0\% polydispersity. In both (a) and (b) there is a strong correlation, which is in contrast to what happens in the case of size polydispersity; thus 
    (c) shows a scatter plot between the single-component system and 30\% size polydispersity. 
    (d) shows how $R$ varies as a function of polydispersity for the two cases of energy polydispersity (blue and green, respectively corresponding to (a) and (b)) and size polydispersity (yellow). In the latter case the correlation coefficient drops quickly.}
    \label{fig7}
\end{figure}

The $3N$-dimensional collective force vector composed of all particle forces is minus the gradient of $U(\bR)$. Two systems with identical $\Omega$ at given state points will have proportional collective force vectors because both are normal to $\Omega$, but the force vectors are not necessarily of the same length. A closer analysis that will not be repeated here \cite{dyr13,dyr13a}, shows that the so-called \textit{reduced} force vectors \cite{IV} are identical. This leads to the same structure and dynamics even though the temperatures may differ for two systems at state points with the same $\Omega$. In the present case of energy polydispersity with energies averaging to unity, there is no need to adjust the temperature, however. Thus one way to confirm the above finding of closely similar $\Omega$s when the energy polydispersity is changed is to look at the distribution of the individual particle forces \cite{ing23}. 

We carried out such an analysis with a focus on calculating the Pearson correlation coefficient, again jumping between 0\% and 30\% polydispersity. \Fig{fig6}(a) shows a scatter plot of the x-coordinates of all particle forces in the 0\% to 30\% case, and (b) shows the reverse change. We find in both cases a strong correlation ($R>0.96$). The linear-regression slopes are close to but not identical to unity. This suggests that a slight adjustment of the temperature would result in an even better collapse of the RDFs and of the intermediate scattering functions (\fig{fig1}), but we have found that this effect is minor and did not investigate it further. For comparison, we show in \fig{fig6}(c) the same analysis when going from 0\% to 30\% size polydispersity. In that case the forces correlate only weakly, and as shown in (d) a significant ``decorrelation'' takes effect even when just a small size polydispersity is introduced (yellow curve). Overall, the picture for the forces is very similar to that of the potential energies (\fig{fig4}).

That the temperatures corresponding to a given $\Omega$ are virtually independent of the energy polydispersity can be validated by considering the so-called configurational temperature $\Tc$, a quantity that refers exclusively to the configurational degrees of freedom \cite{dyr13a,LLstat,rug97,pow05,saw23a,saw23b} with the property that $\Tc=T$ for a system in thermal equilibrium. $\Tc$ is defined as a ratio of two canonical averages, $k_B\Tc\equiv\langle(\nabla U)^2\rangle/\langle \nabla^2 U\rangle$. For a large system $\Tc$ can be reliably evaluated from a single equilibrium configuration because the relative fluctuations of both numerator and denominator go to zero in the thermodynamic limit.

\Fig{fig7} shows plots analogous to those of \fig{fig6} with $\Tc(\bR)\equiv (\nabla U(\bR))^2/ \nabla^2 U(\bR)$ instead of the individual x-components of the particle forces. Note that while the latter gives many data points for each configuration $\bR$, the former yields just a single point, meaning that several configurations are needed to make the plot (as for $U$ in \fig{fig4}). The $\Tc$ results are quite similar to those of \fig{fig6}: Going from 0\% to 30\% energy polydispersity or the reverse, the configurational temperatures correlate strongly, which is not the case for size polydispersity.

\section{Summary}

We have shown that the constant-potential-energy surfaces of energy polydisperse LJ systems are virtually independent of the degree of polydispersity in the range 0\%-30\%, i.e., almost identical to those of the single-component LJ system. By reference to $NVU$ dynamics, this explains why structure and dynamics are not affected by energy polydispersity to any significant degree \cite{shagolsem2015a,sha16,ing23}. This is in contrast to what happens when size polydispersity is introduced, in which case the constant-potential-energy surfaces are not invariant and both structure and dynamics change dramatically \cite{ing23}.

\begin{acknowledgments}
This work was supported by the VILLUM Foundation's \textit{Matter} grant (VIL16515).
\end{acknowledgments}

\newpage
\section*{Appendix}

This Appendix presents results corresponding to those of the main paper using a shifted-force cutoff (A) and studies the effect of randomly introducing polydispersity in four different cases (B). Since the figures confirm the findings of the main paper, few comments are given except for the figure captions.

\subsection{Shifted-force cutoff}
A shifted-force cutoff is defined by adding a constant to the pair force below the cutoff radius in such a way that the pair force at the cutoff radius is zero \cite{tildesley,frenkel}. In $NVE$ simulations, this force continuity results in a much better energy conservation than that of a shifted-potential cutoff. For the single-component LJ system with $\sigma=1$ it has been demonstrated that a shifted-force cutoff at just 1.5 is enough to obtain as accurate structure and dynamics as in simulations of a shifted-potential cutoff at 2.5 \cite{tox11a}, even though the LJ pair force is 30 times larger at distance 1.5 than at distance 2.5. This fact allows for a speed up of simulations of a factor of four at typical liquid state points \cite{tox11a}. Results for the shifted-force-cutoff simulations are given in Fig. 8-13.

\begin{figure}[H]
    \centering
    \includegraphics[width=0.6\linewidth]{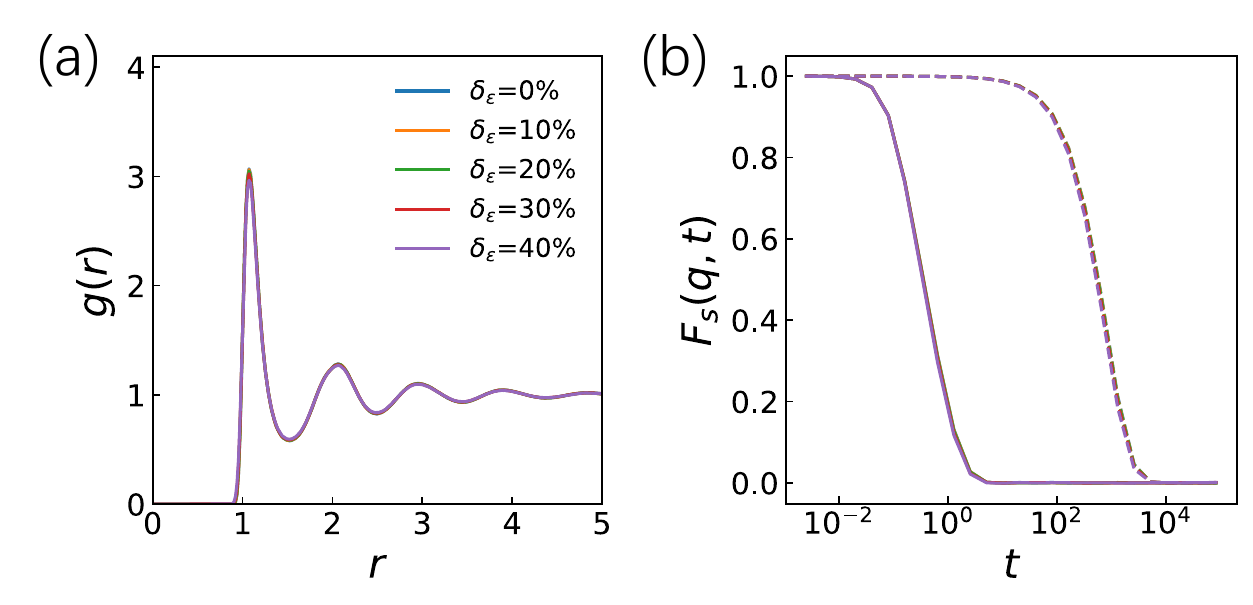}
    \caption{Average structure and dynamic of LJ systems of different energy polydispersity simulated with a shifted-force cutoff at 1.5 at the state point $\left(\rho,T\right) = (0.85,0.70)$; this figure is the shifted-force analog of \fig{fig1}. 
    (a) shows the average RDF for up to 40\% polydispersity. 
    (b) shows the corresponding average incoherent intermediate scattering functions where solid lines represent the wave vector of the first peak of the static structure factor of the single-component LJ system ($q = 7.2$) and dashed lines represent the wave vector corresponding to the box length ($q = 0.19$). For both structure and dynamics, we find as in \fig{fig1} results that are virtually independent of the degree of polydispersity.}
    \label{sfig_RDF&Fs_shiftedforce}
\end{figure}

\begin{figure}[H]
    \centering
    \includegraphics[width=1.0\linewidth]{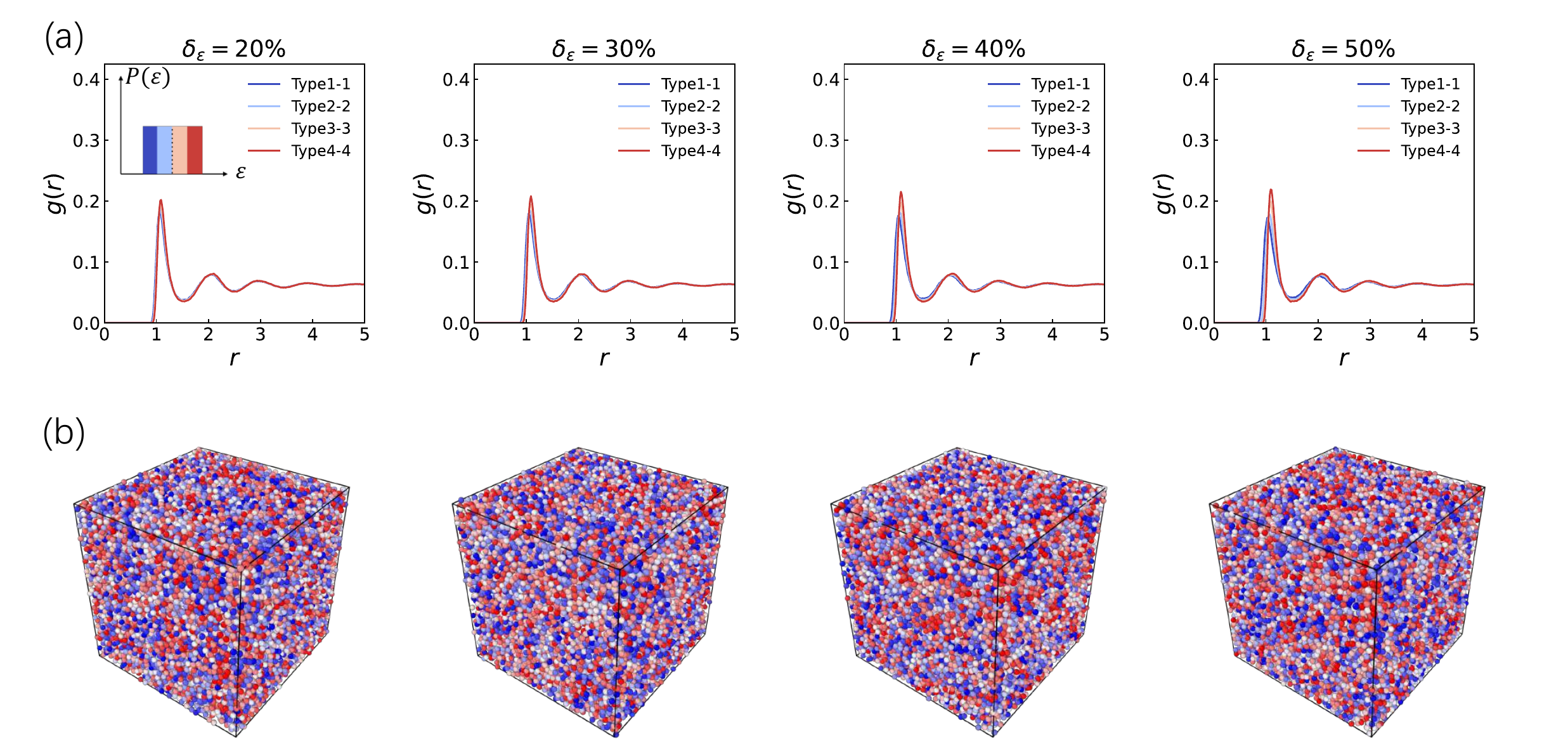}
    \caption{Structure of LJ systems with energy polydispersity varying between 20\% and 50\%; this figure is the shifted-force analog of \fig{fig2}. 
    (a) The particles are divided into four categories according to their energy as illustrated in the inset of the first panel. With increasing polydispersity the energy-resolved RDFs differ more and more. Interestingly, the difference is less pronounced than for the shifted-potential simulations of the main paper.
    (b) shows snapshots for each polydispersity in which the red particles have large energy and the blue have small energy. In contrast to the findings of the main paper, there is no clearly visible phase separation at the highest polydispersity.}
    \label{sfig_RdfperType_shiftedforce}
\end{figure}

\begin{figure}[H]
    \centering
    \includegraphics[width=0.8\linewidth]{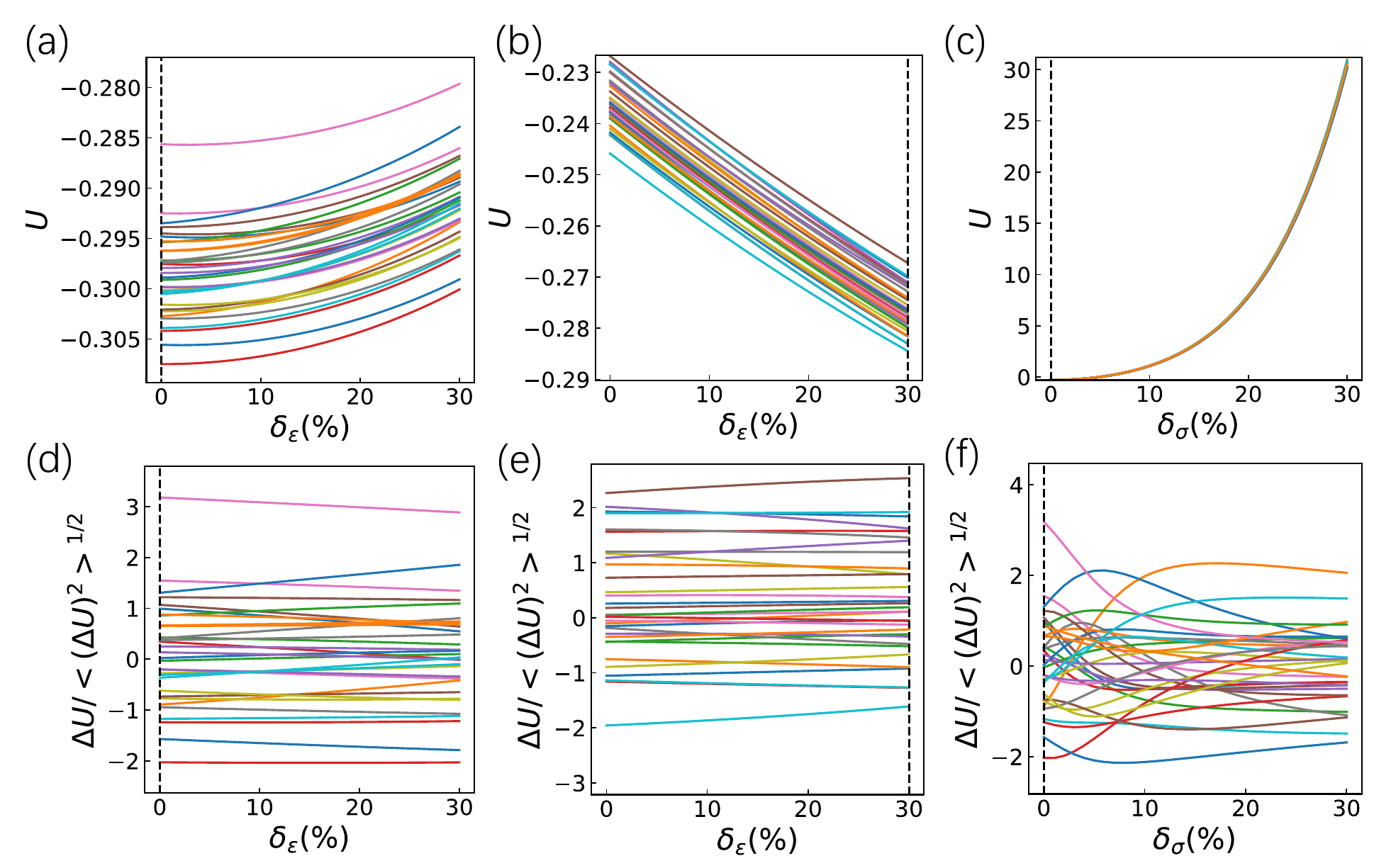}
    \caption{Potential energies of configurations as a function of the degree of polydispersity; this figure is the shifted-force analog of \fig{fig3}. Note that the potential energy used here differs from that of the main paper, reflecting the fact that the pair forces are modified below the cutoff.
    (a) and (b) show the potential energies of 32 independent configurations taken from an equilibrium simulation at 0\% and 30\% energy polydispersity, respectively. A closer look at the level crossings is provided in (d) and (e) that show the relative potential-energy variations of the data in (a) and (b). The observed rare crossings mean that energy polydisperse LJ systems to a good approximation conform to \eq{crit}, which is equivalent to having invariant $\Omega$. For comparison, (c) and (f) show the effect of introducing size polydispersity.}
    \label{sfig_U_shiftedforce}
\end{figure} 

\begin{figure}[H]
    \centering
    \includegraphics[width=\linewidth]{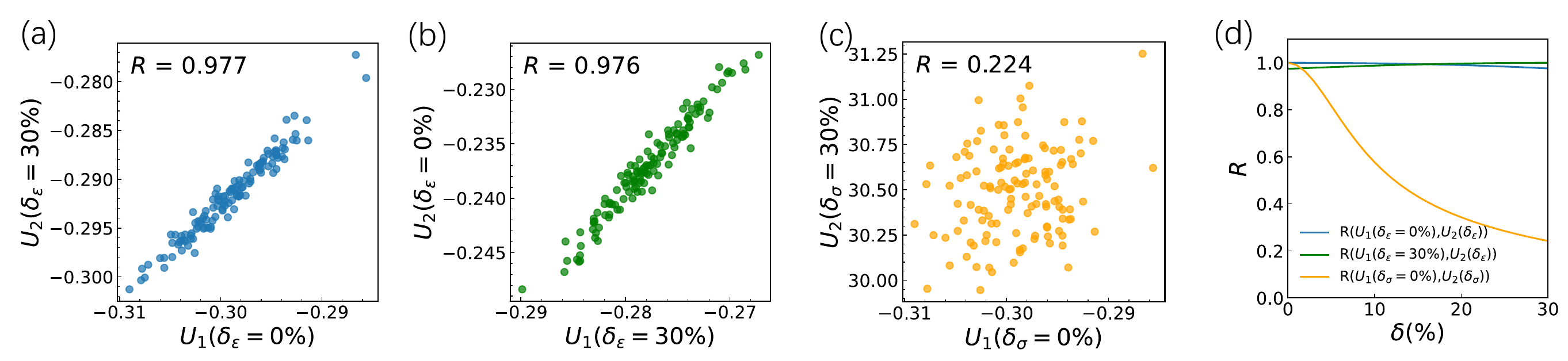}
    \caption{Correlation between the potential energy of configurations sampled from the equilibrium state of the potential $U_1$ corresponding to some polydispersity, subsequently evaluated with respect to a different polydispersity corresponding to the potential energy $U_2$; this figure is the shifted-force analog of \fig{fig4}.
    (a) $R$ between the  single-component system, $U_1$, and 30\% energy polydispersity, $U_2$. 
    (b) The reverse of (a); configurations were here equilibrated at 30\% energy polydispersity and subsequently evaluated with respect to zero polydispersity. In both (a) and (b) there is a strong correlation. The strong correlations observed in (a) and (b) are in contrast to what happens in the case of size polydispersity for which 
    (c) shows $R$ between the single-component system 30\% size polydispersity. 
    (d) shows how $R$ varies as a function of polydispersity for the two cases of energy polydispersity (blue and green) and size polydispersity (yellow). In the latter case the correlation drops quickly.}
    \label{sfig_RforU_shiftedforce}
\end{figure}

\begin{figure}[H]
    \centering
    \includegraphics[width=\linewidth]{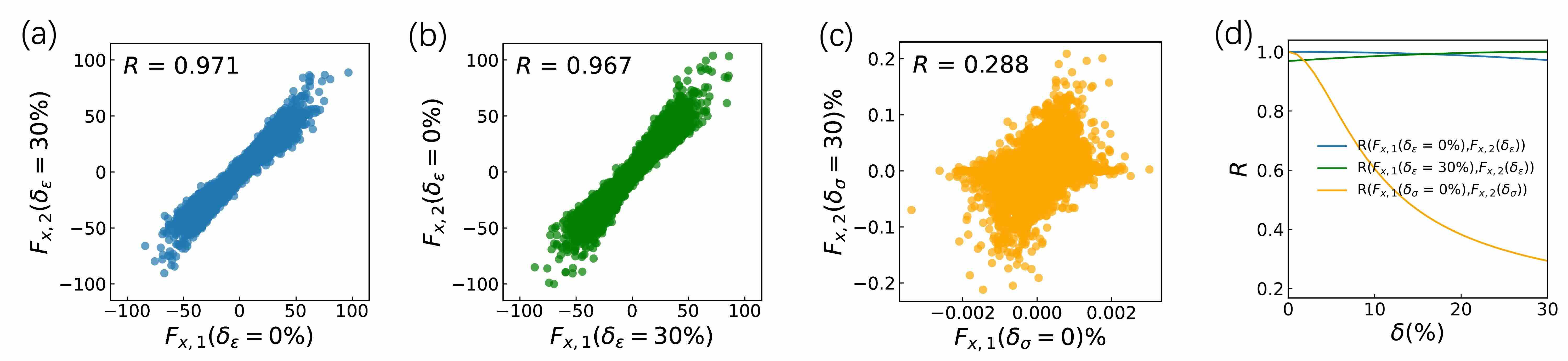}
    \caption{Correlations between the individual particle forces in the x-direction, $F_x$, of a single configuration sampled from the equilibrium state of some polydispersity, subsequently evaluated with respect to a different polydispersity; this figure is the shifted-force analog of \fig{fig6}.
    (a) Correlation between the single-component system and 30\% energy polydispersity. 
    (b) The reverse of (a); the configuration is here an equilibrium configuration at 30\% energy polydispersity and the forces were subsequently evaluated with respect to 0\% polydispersity. In both (a) and (b) there is a strong correlation, which is in contrast to what happens in the case of size polydispersity; thus 
    (c) shows the correlation between the single-component system and 30\% size polydispersity. 
    (d) shows how $R$ varies as a function of polydispersity for the two cases of energy polydispersity (blue and green) and size polydispersity (yellow). In the latter case the correlation coefficient drops quickly.}
    \label{sfig_RforFx_shiftedforce}
\end{figure}

\begin{figure}[H]
    \centering
    \includegraphics[width=\linewidth]{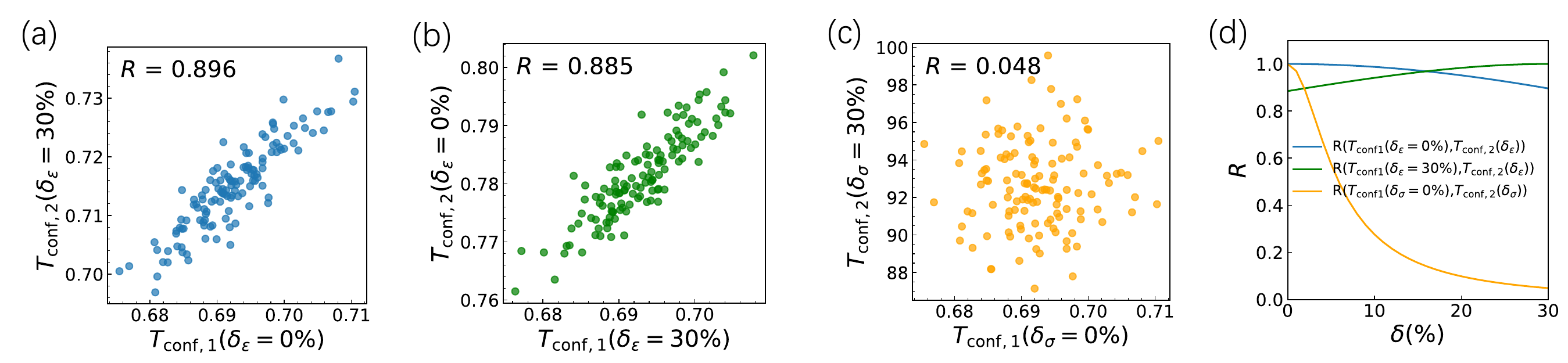}
    \caption{Correlations between the configurational temperatures, $\Tc$, of configurations sampled from the equilibrium state of some polydispersity and the configurational temperature evaluated with respect to a different polydispersity; this figure is the shifted-force analog of \fig{fig7}.
    (a) Correlation between the single-component system and 30\% energy polydispersity.    
    (b) The reverse of (a); configurations were here equilibrated at 30\% energy polydispersity and the configurational temperatures subsequently evaluated with respect to 0\% polydispersity. In both (a) and (b) there is a strong correlation, which is in contrast to what happens in the case of size polydispersity; thus 
    (c) shows a scatter plot between the single-component system and 30\% size polydispersity. 
    (d) shows how $R$ varies as a function of polydispersity for the two cases of energy polydispersity (blue and green) and size polydispersity (yellow). In the latter case the correlation coefficient drops quickly.}
    \label{sfig_RforTconf_shiftedforce}
\end{figure}

\subsection{Other choices of random energies}

Figures 14-17 demonstrate that the approximate invariance of $\Omega$ with changing energy polydispersity does not depend on the choice of the random energies.

\begin{figure}[H]
    \centering
    \includegraphics[width=\linewidth]{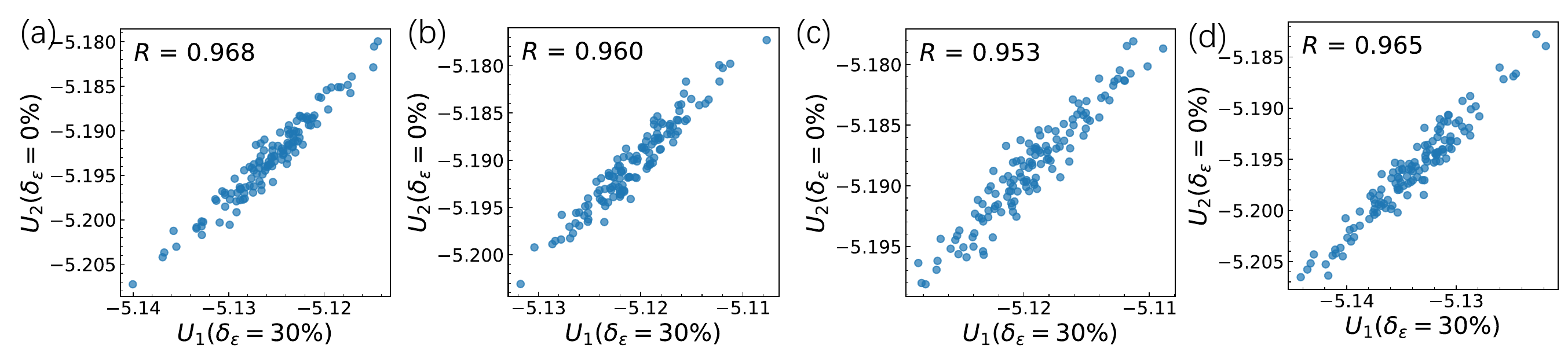}
    \caption{Potential energies of configurations as a function of the degree of energy polydispersity for four different choices of random energies. In each case 32 independent configurations are taken from an equilibrium simulation at 0\% energy polydispersity.}
    \label{sfig_U_otherrandomseed}
\end{figure}

\begin{figure}[H]
    \centering
    \includegraphics[width=\linewidth]{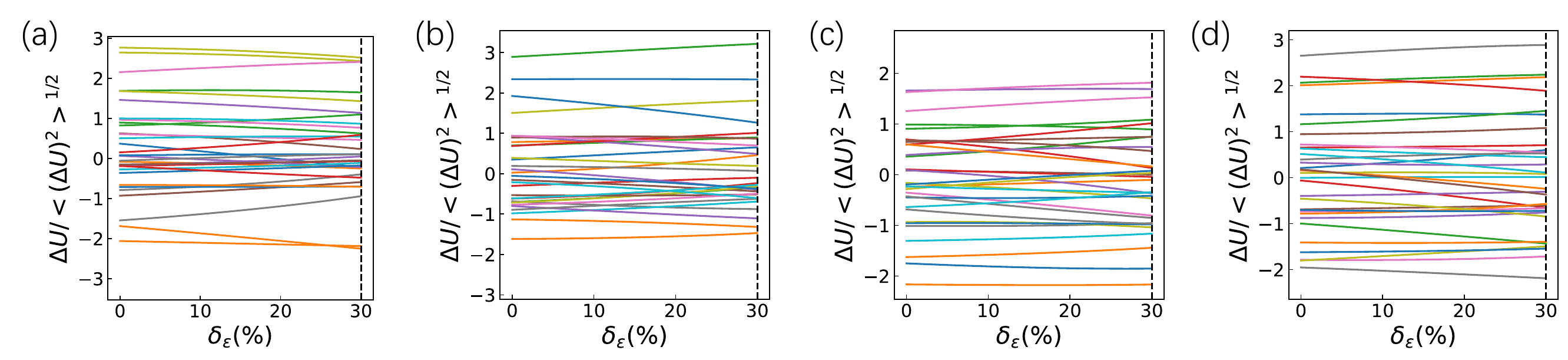}
    \caption{Potential energies of configurations as a function of the degree of energy polydispersity with four different choices of random energies. In each case 32 independent configurations are taken from an equilibrium simulation at 30\% energy polydispersity. The 32 configurations sampled in each subplot are different since we choose different choices of random energies at the beginning of the 30\% polydispersity equilibrium simulation, while in Fig. \ref{sfig_RforU_otherrandomseed} below the configurations are the same. The figure demonstrates that \eq{crit} is obeyed independent of the choice of random energies.}
    \label{sfig_U_otherrandomseed_30to0}
\end{figure}

\begin{figure}[H]
    \centering
    \includegraphics[width=\linewidth]{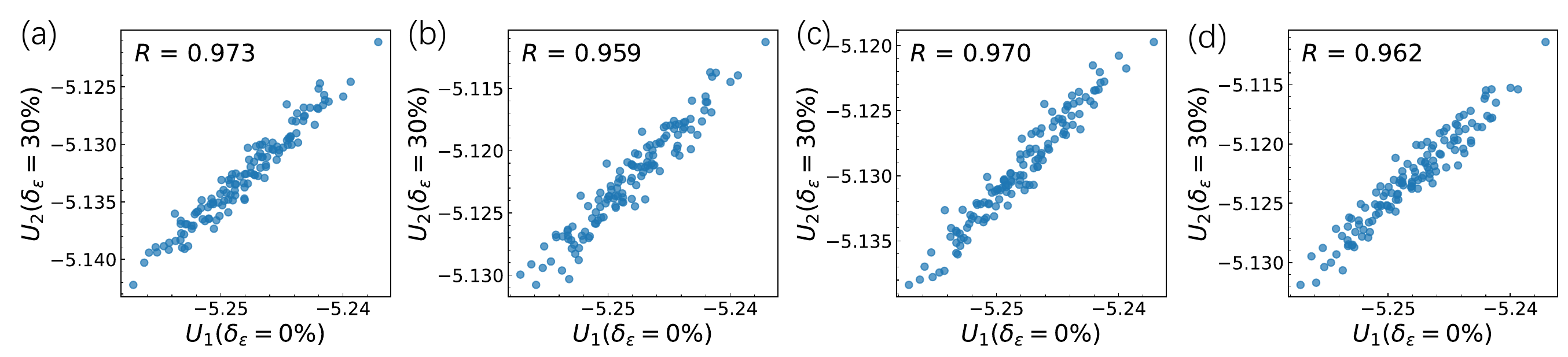}
    \caption{Correlations between the potential energy of configurations sampled from the equilibrium state of the single-component system $U_1$, subsequently evaluated with respect to 30\% energy polydispersity $U_2$, with four different choices of random energies. }
    \label{sfig_RforU_otherrandomseed}
\end{figure}

\begin{figure}[H]
    \centering
    \includegraphics[width=\linewidth]{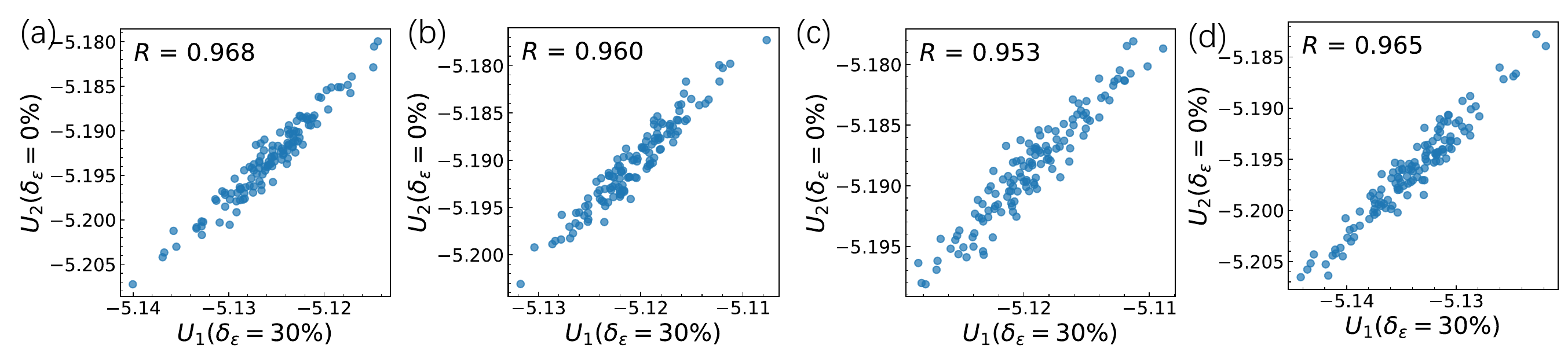}
    \caption{Correlations between the potential energy of configurations sampled from the equilibrium
state of 30\% energy polydispersity system $U_1$, subsequently evaluated with respect
to zero polydispersity, with four different choices of random energies.}
    \label{sfig_RforU_otherrandomseed_30to0}
\end{figure}

\end{document}